\documentclass[twocolumn,pre]{revtex4-1}
\usepackage{mathrsfs}
\usepackage{bm}
\usepackage{amsmath}
\usepackage{amssymb}
\usepackage{graphicx}
\usepackage{esint}

\usepackage{hyperref}
\usepackage{color}

\definecolor{myblue}{RGB}{0,0,255}
\hypersetup{
    colorlinks,
    citecolor=myblue,
    linkcolor=myblue,
    urlcolor=myblue
}

\begin{document}
\title{Uncertainty relations in stochastic processes: An information inequality
approach}
\author{Yoshihiko Hasegawa}
\email{hasegawa@biom.t.u-tokyo.ac.jp}
\affiliation{Department of Information and Communication Engineering, Graduate
School of Information Science and Technology, The University of Tokyo,
Tokyo 113-8656, Japan}
\author{Tan Van Vu}
\email{tan@biom.t.u-tokyo.ac.jp}
\affiliation{Department of Information and Communication Engineering, Graduate
School of Information Science and Technology, The University of Tokyo,
Tokyo 113-8656, Japan}
\begin{abstract}
The thermodynamic uncertainty relation is an inequality stating that
it is impossible to attain higher precision than the bound defined
by entropy production. In statistical inference theory, information
inequalities assert that it is infeasible for any estimator to achieve
an error smaller than the prescribed bound. Inspired by the similarity
between the thermodynamic uncertainty relation and the information
inequalities, we apply the latter to systems described by Langevin
equations and derive the bound for the fluctuation of thermodynamic
quantities. When applying the Cram\'er--Rao inequality, the obtained
inequality reduces to the fluctuation-response inequality. We find
that the thermodynamic uncertainty relation is a particular case of
the Cram\'er--Rao inequality, in which the Fisher information is
the total entropy production. Using the equality condition of the
Cram\'er--Rao inequality, we find that the stochastic total entropy
production is the only quantity which can attain equality in the
thermodynamic uncertainty relation. Furthermore, we apply the Chapman--Robbins
inequality and obtain a relation for the lower bound of the
ratio between the variance and the sensitivity of systems in response
to arbitrary perturbations.

\end{abstract}
\maketitle

\section{Introduction}

Over the last two decades, substantial progress has been made in terms
of universal relations among thermodynamic quantities, such as fluctuation
theorems and generalized second laws \cite{Ritort:NEArticle,Seifert:2012:FTReview,VandenBroeck:2015:Review,Parrondo:2015:InfoThermo}.
One of the key achievements in this area is the thermodynamic uncertainty
relation \cite{Barato:2015:UncRel,Barato:2015:FanoBound,Gingrich:2016:TUP,Polettini:2016:TUP,Pietzonka:2016:Bound,Horowitz:2017:TUR,Proesmans:2017:TUR,Pietzonka:2017:FiniteTUR,Pigolotti:2017:EP,Dechant:2018:TUR},
which states that fluctuations in thermodynamic quantities are bounded
from below by the reciprocal of entropy production. 
The thermodynamic uncertainty relation provides the theoretical justification for our intuition that
higher precision is inevitably accompanied with larger energy consumption.
Universal relations
between ``cost'' and ``quality'' also exist in fields other than
thermodynamics. It is an empirical truism that as the amount of available
data increases, inferences on parameters become more precise. Information
inequalities provide theoretical support for this intuition by giving the lower bounds for estimators. Information inequalities are
known to be the basis for inequalities in other fields; for instance,
Heisenberg's uncertainty principle can be derived through these inequalities
\cite{Dembo:1990:InfoUnc,Dembo:1991:InfoIneq}. The universality
of information inequalities leads us to posit that they may play
an important role in stochastic thermodynamic systems.

Herein, we regard fluctuations of thermodynamic quantities as
errors in statistical estimators, thereby obtaining inequality relations
for quantities of stochastic processes. In particular, we obtain the
Cram\'er--Rao inequality for systems described by Langevin equations
which relates the fluctuation of thermodynamic quantities to the
Fisher information. This relation reduces to a recently discovered
fluctuation-response inequality \cite{Dechant:2018:FRI}. We
show that the thermodynamic uncertainty relation is a particular case of the Cram\'er--Rao
inequality in which the Fisher information is the total entropy production.
Using the equality condition of the Cram\'er--Rao inequality, we
find that the stochastic total entropy production is the only quantity
which can attain equality in the thermodynamic uncertainty relation. Furthermore, we apply the
Chapman--Robbins inequality, which is a generalization of the Cram\'er--Rao
inequality, to the systems to show that the ratio between the variance
and the sensitivity is bounded from below by the reciprocal of the
Pearson divergence for any perturbation. As an application of the
Chapman--Robbins inequality, we obtain an explicit inequality between
the phase variance and the phase sensitivity of stochastic limit cycle
oscillators.

\section{Model}

We consider the following $N$-dimensional Ito Langevin equation for
$\bm{x}\equiv[x_{1},x_{2},...,x_{N}]^{\top}$: 
\begin{align}
\dot{\bm{x}} & =\bm{A}_{\theta}(\bm{x},t)+\sqrt{2}\bm{C}(\bm{x},t)\bm{\xi}(t),\label{eq:dynamics_def}
\end{align}
where $\bm{\xi}(t)\equiv[\xi_{1}(t),...,\xi_{M}(t)]^{\top}$ is
white Gaussian noise with $\left\langle \xi_{i}(t)\xi_{j}(t')\right\rangle =\delta_{ij}\delta(t-t')$
($M$ is the number of noise terms), $\bm{A}_{\theta}(\bm{x},t)\equiv[A_{\theta,1}(\bm{x},t),...,A_{\theta,N}(\bm{x},t)]^{\top}$
is a drift vector with a real parameter $\theta$, and $\bm{C}(\bm{x},t)\equiv[C_{ij}(\bm{x},t)]$
is an $N\times M$ noise matrix. $\theta$ is a parameter to be estimated
with pre-defined estimators. For simplicity, we assume that $\theta$
is a scalar, but the calculation can be easily generalized to a multidimensional
$\bm{\theta}$. Let $P_{\theta}(\bm{x},t)$ be the probability density
function of $\bm{x}$ at time $t$. Defining $[B_{ij}(\bm{x},t)]=\bm{B}(\bm{x},t)\equiv\bm{C}(\bm{x},t)\bm{C}(\bm{x},t)^{\top}$,
the Fokker--Planck equation (FPE) of Eq.~\eqref{eq:dynamics_def}
is \cite{Risken:1989:FPEBook,Gardiner:2009:Book} 
\begin{equation}
\partial_{t}P_{\theta}(\bm{x},t)=\widehat{\mathcal{L}}_{\theta}(\bm{x},t)P_{\theta}(\bm{x},t),\label{eq:FPE_def}
\end{equation}
 where $\widehat{\mathcal{L}}_{\theta}(\bm{x},t)\equiv-\sum_{i}\partial_{x_{i}}A_{\theta,i}(\bm{x},t)+\sum_{i,j}\partial_{x_{i}}\partial_{x_{j}}B_{ij}(\bm{x},t)$
is an FPE operator. The probability current is 
\begin{equation}
J_{\theta,i}(\bm{x},t)\equiv\left\{ A_{\theta,i}(\bm{x},t)-\sum_{j}\partial_{x_{j}}B_{ij}(\bm{x},t)\right\} P_{\theta}(\bm{x},t).\label{eq:J_theta_i_def}
\end{equation}

Now we consider the estimation of the parameter $\theta$ from the measurement
of a stochastic trajectory generated by Eq.~(\ref{eq:dynamics_def})
over an interval from $t=0$ to $T$. Let $\bm{\Gamma}\equiv\left[\bm{x}(t)\right]_{t=0}^{t=T}$
be the measured trajectory and $\mathcal{P}_{\theta}(\bm{\Gamma})$
be the probability of $\bm{\Gamma}$  (Fig.~\ref{fig:FIG1}(a)). 
For an arbitrary function $f(\bm{\Gamma})$,
we define its expectation as $\left\langle f(\bm{\Gamma})\right\rangle _{\theta}\equiv\int\mathcal{D}\bm{\Gamma}\,f(\bm{\Gamma})\mathcal{P}_{\theta}(\bm{\Gamma})$.
We consider an estimator $\Theta(\bm{\Gamma})$ which is an unbiased
estimator for $\psi(\theta)$ and thus we have $\left\langle \Theta(\bm{\Gamma})\right\rangle _{\theta}=\psi(\theta)$.
Since $\bm{\Gamma}$ is a stochastic trajectory, $\Theta(\bm{\Gamma})$ is a random
variable. Therefore, if we repeat the measurement and the estimation,
$\Theta(\bm{\Gamma})$ is distributed around $\psi(\theta)$,
whose variance is identified as the variance of the estimator $\Theta(\bm{\Gamma})$ (Fig.~\ref{fig:FIG1}(a)). 

\section{Cram\'er--Rao inequality}

\subsection{Derivation of uncertainty relation}

The Cram\'er--Rao inequality provides the lower
bound for the variance of estimators. Applying the Cram\'er--Rao
inequality \cite{Kendall:ClassicalInference:2A,George:2001:StatInfer,Lehmann:2003:EstimationBook}
to $\Theta(\bm{\Gamma})$, the following relation holds (Appendix~\ref{sec:info_ineqs}):
\begin{alignat}{1}
\frac{\mathrm{Var}_{\theta}\left[\Theta(\bm{\Gamma})\right]}{\left({\displaystyle \partial_{\theta}\left\langle \Theta(\bm{\Gamma})\right\rangle _{\theta}}\right)^{2}} & \ge\frac{1}{\mathcal{I}(\theta)},\label{eq:TCRI}
\end{alignat}
where $\mathrm{Var}_{\theta}\left[f(\bm{\Gamma})\right]\equiv\left\langle \left\{ f(\bm{\Gamma})-\left\langle f(\bm{\Gamma})\right\rangle _{\theta}\right\} ^{2}\right\rangle _{\theta}$
and $\mathcal{I}(\theta)$ is the Fisher information:
\begin{equation}
\mathcal{I}(\theta)\equiv\left\langle \left(\frac{\partial}{\partial\theta}\ln\mathcal{P}_{\theta}(\bm{\Gamma})\right)^{2}\right\rangle _{\theta}=-\left\langle \frac{\partial^{2}}{\partial\theta^{2}}\ln\mathcal{P}_{\theta}(\bm{\Gamma})\right\rangle _{\theta}.\label{eq:Fisher_I_def}
\end{equation}
 Let $\mathcal{P}_{\theta}(\bm{\Gamma}|\bm{x}^{0})$ be the probability
of $\bm{\Gamma}$ given $\bm{x}^{0}$ at $t=0$. By using a path integral
\cite{Chernyak:2006:PathIntegral,Bressloff:2014:WKB,Wio:2013:PIBook,Dechant:2018:TUR}, 
we obtain (Appendix~\ref{sec:path_integral})
\begin{align}
\mathcal{P}_{\theta}(\bm{\Gamma}|\bm{x}^{0}) & =\mathscr{N}\exp\left[-\int_{0}^{T}dt\,\mathscr{A}_{\theta}(\bm{x}(t),t)\right],\label{eq:PI_def}\\
\mathscr{A}_{\theta}(\bm{x},t) & \equiv\frac{1}{4}\left\{ \left(\dot{\bm{x}}-\bm{A}_{\theta}\right)^{\top}\bm{B}^{-1}\left(\dot{\bm{x}}-\bm{A}_{\theta}\right)\right\} ,\label{eq:stoc_action_A_def}
\end{align}
where $\mathscr{N}$ is a term that does not depend on $\theta$.
In Eq.~\eqref{eq:PI_def}, we employ a pre-point discretization.
Due to the pre-point discretization,
cross terms such as $A_{\theta}(x,t)B(x,t)^{-1}\dot{x}$ should
be interpreted as $A_{\theta}(x,t)B(x,t)^{-1}\bullet\dot{x}$,
where $\bullet$ denotes the Ito product. 
Although $\mathscr{A}_{\theta}(\bm{x},t)$ is different for a mid-point
discretization \cite{Tang:2014:PI}, both discretizations reduce
to the same expression for additive noise systems \cite{Adib:2008:PI}.
We have $\mathcal{P}_{\theta}(\bm{\Gamma})=\mathcal{P}_{\theta}(\bm{\Gamma}|\bm{x}^{0})P_{\theta}(\bm{x}^{0})$
where $P_{\theta}(\bm{x}^{0})$ is the initial probability density
of $\bm{x}^{0}$ at $t=0$ ($\int\mathcal{D}\bm{\Gamma}\mathcal{P}_{\theta}(\bm{\Gamma})=1$).
The log-probability is calculated as 
\begin{align}
\ln\mathcal{P}_{\theta}(\bm{\Gamma}) & =\ln\mathscr{N}+\ln P_{\theta}(\bm{x}^{0})\nonumber \\
 & -\frac{1}{4}\int_{0}^{T}dt\,\left(\dot{\bm{x}}-\bm{A}_{\theta}\right)^{\top}\bm{B}^{-1}\left(\dot{\bm{x}}-\bm{A}_{\theta}\right).\label{eq:loglikelihood_def}
\end{align}
From Eq.~\eqref{eq:Fisher_I_def}, we need to calculate the second
derivative of Eq.~\eqref{eq:loglikelihood_def} with respect to $\theta$:
\begin{align}
 & \frac{\partial^{2}}{\partial\theta^{2}}\ln\mathcal{P}_{\theta}(\bm{\Gamma})\nonumber \\
 & =\frac{\partial^{2}}{\partial\theta^{2}}\ln P_{\theta}(\bm{x}^{0})-\frac{1}{2}\int_{0}^{T}dt\,\left(\frac{\partial}{\partial\theta}\bm{A}_{\theta}\right)^{\top}\bm{B}^{-1}\left(\frac{\partial}{\partial\theta}\bm{A}_{\theta}\right)\nonumber \\
 & +\frac{1}{2}\int_{0}^{T}dt\,\left(\dot{\bm{x}}-\bm{A}_{\theta}\right)^{\top}\bullet\bm{B}^{-1}\left(\frac{\partial^{2}}{\partial\theta^{2}}\bm{A}_{\theta}\right).\label{eq:second_derivative}
\end{align}
When applying the expectation
$\left\langle \cdots\right\rangle _{\theta}$ to Eq.~\eqref{eq:second_derivative},
the last term disappears (cf. Eq.~\eqref{eq:non_anticipating_vanish}
in Appendix~\ref{sec:path_integral}). Therefore, from Eq.~\eqref{eq:Fisher_I_def},
the Fisher information is given by
\begin{align}
\mathcal{I}(\theta) & =-\left\langle \frac{\partial^{2}}{\partial\theta^{2}}\ln P_{\theta}(\bm{x}^{0})\right\rangle _{\theta}\nonumber \\
 & +\frac{1}{2}\left\langle \int_{0}^{T}dt\left(\frac{\partial}{\partial\theta}\bm{A}_{\theta}^{\top}\right)\bm{B}^{-1}\left(\frac{\partial}{\partial\theta}\bm{A}_{\theta}\right)\right\rangle _{\theta}.\label{eq:FI_TCRI}
\end{align}

Equation~\eqref{eq:TCRI} reduces to the
recently proposed fluctuation-response inequality \cite{Dechant:2018:FRI}. Suppose $\bm{A}_{\theta}(\bm{x},t)=\bm{\mathcal{A}}(\bm{x},t)+\theta\bm{\mathcal{Z}}(\bm{x},t)$,
where $\theta$ is a sufficiently small real parameter and $\bm{\mathcal{Z}}(\bm{x},t)$
is an arbitrary perturbation. Since $\theta$ is sufficiently small,
${\displaystyle \partial_{\theta}\left\langle \Theta(\bm{\Gamma})\right\rangle _{\theta}}$
can be approximated by $\left.\partial_{\theta}\left\langle \Theta(\bm{\Gamma})\right\rangle _{\theta}\right|_{\theta=0}\simeq\left(\left\langle \Theta(\bm{\Gamma})\right\rangle _{\theta}-\left\langle \Theta(\bm{\Gamma})\right\rangle _{\theta=0}\right)/\theta$.
Entering these expressions into Eq.~\eqref{eq:TCRI}, we obtain the
fluctuation-response inequality: 
\begin{equation}
\frac{\mathrm{Var}_{\theta=0}\left[\Theta(\bm{\Gamma})\right]}{\left[\left\langle \Theta(\bm{\Gamma})\right\rangle _{\theta}-\left\langle \Theta(\bm{\Gamma})\right\rangle _{\theta=0}\right]^{2}}\ge\frac{1}{\mathcal{C}}=\frac{1}{\theta^{2}\mathcal{I}(0)},\label{eq:FRI_def}
\end{equation}
where $\mathcal{C}\equiv\theta^{2}\mathcal{I}(0)$ and $\mathcal{I}(0)=\frac{1}{2}\left\langle \int_{0}^{T}dt\,\bm{\mathcal{Z}}^{\top}\bm{B}^{-1}\bm{\mathcal{Z}}\right\rangle _{\theta=0}$.
Note that the boundary term is ignored in Eq.~\eqref{eq:FRI_def}. 
The boundary term is zero when the probability density function
remains unchanged upon perturbation, which is the case for
the thermodynamic uncertainty relation. 
From Eq.~\eqref{eq:FRI_def}, $1/\left[\theta^{2}\mathcal{I}(0)\right]$
is the lower bound of the fluctuation-response inequality.
Because the fluctuation-response inequality holds only for sufficiently small $\theta$, 
the perturbation should be sufficiently weak. 

\begin{figure}
\includegraphics[width=8.5cm]{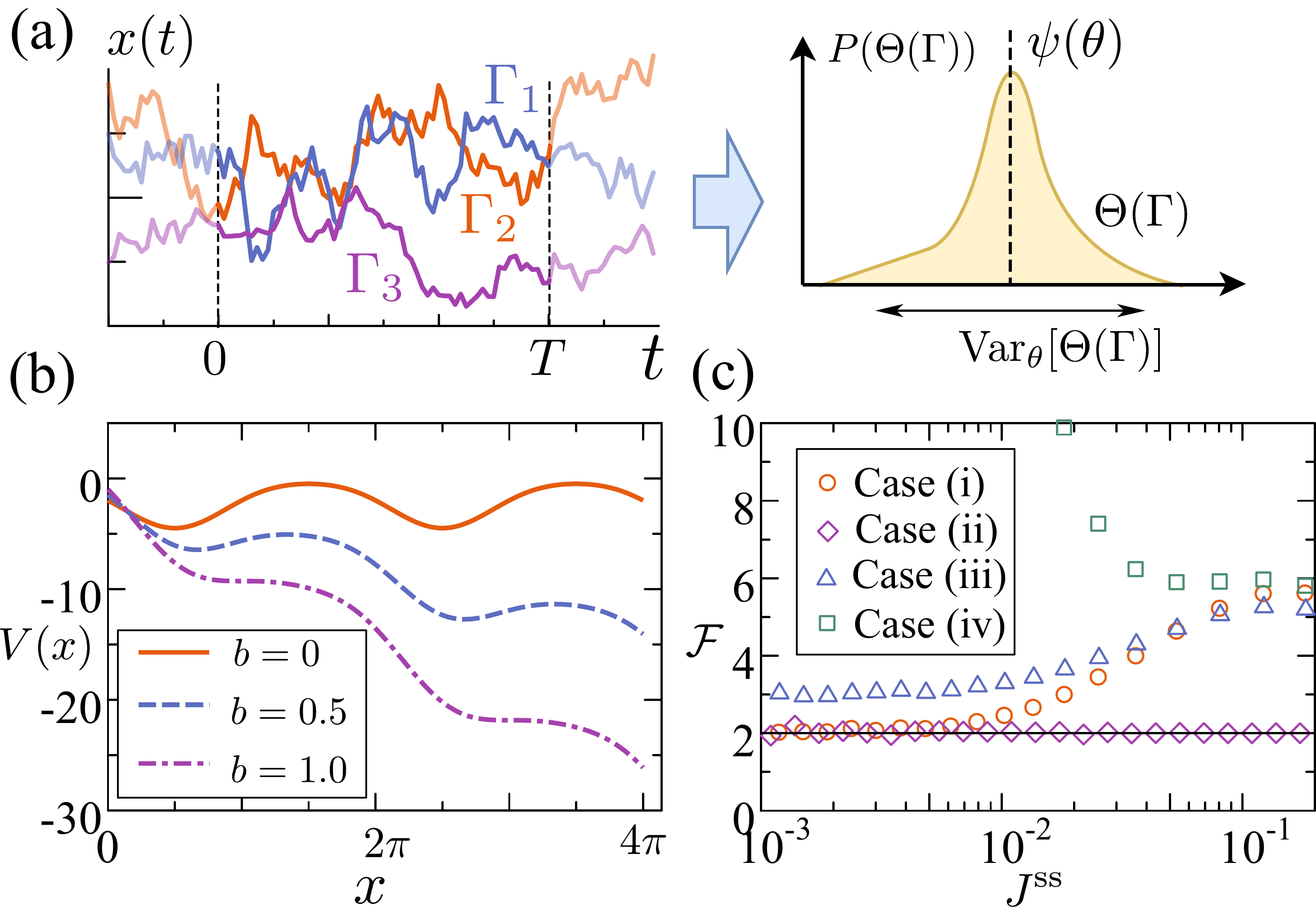}

\caption{(a) Relation between trajectory $\Gamma$ and its estimator $\Theta(\Gamma)$.
Given a trajectory $\Gamma=[x(t)]_{t=0}^{t=T}$, $\Theta(\Gamma)$
is an unbiased estimator for $\psi(\theta)$.
Repeating the measurement and estimation, we obtain the 
probability density of $\Theta(\Gamma)$. 
The variance of the probability density corresponds to $\mathrm{Var}_{\theta}[\Theta(\Gamma)]$,
where $\langle \Theta(\Gamma) \rangle_\theta = \psi(\theta)$. 
(b) Effective potential function considered in the numerical verification. The effective
potential $V(x)$ is plotted for $b=0$ (solid line), $0.5$ (dashed
line), and $1.0$ (dot-dashed line) with $a=2$. (c) Numerical verification
of the equality condition of the thermodynamic uncertainty relation. $\mathcal{F}$ are plotted as
a function of $J^{\mathrm{ss}}$, where $\mathcal{F}=2$ for the equality
of the thermodynamic uncertainty relation. Cases (i) (circles) and (ii) (diamonds) satisfy the equality
condition while cases (iii) (triangles) and (iv) (squares) do not.
The parameter settings are $D=1.0$, $a=2$, $b\in[0.001,1.0]$ and $T=4.0$.
\label{fig:FIG1}}
\end{figure}
We explore the thermodynamic uncertainty relation from a statistical inference perspective. Reference~\cite{Dechant:2018:FRI}
provides an alternative derivation of
 the finite-time current thermodynamic uncertainty relation \cite{Horowitz:2017:TUR}
with the fluctuation-response inequality using the notion of a virtual perturbation. Let us consider
a system $\bm{A}(\bm{x},t)=\bm{A}(\bm{x})$ and $\bm{C}(\bm{x},t)=\bm{C}(\bm{x})$
in Eq.~\eqref{eq:dynamics_def} at a steady state. The thermodynamic uncertainty relation considers
the generalized current 
\begin{equation}
\Theta_{\mathrm{cur}}(\bm{\Gamma})\equiv\int_{0}^{T}\bm{\Lambda}(\bm{x})^{\top}\circ\dot{\bm{x}}dt,\label{eq:Thet_cur_def}
\end{equation}
where $\bm{\Lambda}(\bm{x})\equiv[\Lambda_{1}(\bm{x}),...,\Lambda_{N}(\bm{x})]^{\top}$
is an arbitrary projection function and $\circ$ denotes the Stratonovich
product. Using the virtual perturbation technique \cite{Dechant:2018:FRI},
we define 
\begin{equation}
A_{\theta,i}(\bm{x})\equiv(\theta+1)A_{i}(\bm{x})-\frac{\theta}{P^{\mathrm{ss}}(\bm{x})}\sum_{j}\partial_{x_{j}}B_{ij}(\bm{x})P^{\mathrm{ss}}(\bm{x}),\label{eq:A_mod_def}
\end{equation}
where $P^{\mathrm{ss}}(\bm{x})$ is the steady-state distribution
of the unperturbed dynamics (i.e., the dynamics for the case where $\theta=0$).
Note that the steady-state distribution corresponding to $\bm{A}_{\theta}(\bm{x})$ of Eq.~\eqref{eq:A_mod_def}
does not depend on $\theta$ \cite{Dechant:2018:FRI}. 
Using
Eq.~\eqref{eq:A_mod_def}, we find 
\begin{align}
\left\langle \Theta_{\mathrm{cur}}(\bm{\Gamma})\right\rangle _{\theta} & =\left\langle \int_{0}^{T}\bm{\Lambda}(\bm{x})^{\top}\circ\dot{\bm{x}}dt\right\rangle _{\theta}\nonumber \\
 & =T\int d\bm{x}\,\bm{\Lambda}(\bm{x})^{\top}\bm{J}_{\theta}^{\mathrm{ss}}(\bm{x})\nonumber \\
 & =T\int d\bm{x}\,\bm{\Lambda}(\bm{x})^{\top}(1+\theta)\bm{J}^{\mathrm{ss}}(\bm{x})\nonumber \\
 & =(\theta+1)\jmath,\label{eq:Theta_cur_expec_def}
\end{align}
where $\bm{J}^{\mathrm{ss}}(\bm{x})\equiv[J_{1}^{\mathrm{ss}}(\bm{x}),...,J_{N}^{\mathrm{ss}}(\bm{x})]^{\top}$
and $\bm{J}_{\theta}^{\mathrm{ss}}(\bm{x})\equiv[J_{\theta,1}^{\mathrm{ss}}(\bm{x}),...,J_{\theta,N}^{\mathrm{ss}}(\bm{x})]^{\top}$
are the steady-state probability currents of unperturbed and perturbed
dynamics, respectively, and $\jmath\equiv\left\langle \Theta_{\mathrm{cur}}(\bm{\Gamma})\right\rangle _{\theta=0}=T\int d\bm{x}\,\bm{\Lambda}(\bm{x})^{\top}\bm{J}^{\mathrm{ss}}(\bm{x})$.
$\jmath$ corresponds to the averaged current and $\partial_{\theta}\left\langle \Theta_{\mathrm{cur}}(\bm{\Gamma})\right\rangle _{\theta}=\jmath$
from Eq.~\eqref{eq:Theta_cur_expec_def}. From Eq.~\eqref{eq:FI_TCRI},
the Fisher information is
\begin{align}
\mathcal{I}(0) & =\frac{1}{2}\left\langle \int_{0}^{T}dt\left(\frac{\bm{J}^{\mathrm{ss}}(\bm{x})^{\top}}{P^{\mathrm{ss}}(\bm{x})}\right)\bm{B}(\bm{x})^{-1}\left(\frac{\bm{J}^{\mathrm{ss}}(\bm{x})}{P^{\mathrm{ss}}(\bm{x})}\right)\right\rangle _{\theta=0}\nonumber \\
 & =\frac{T}{2}\int d\bm{x}\frac{\bm{J}^{\mathrm{ss}}(\bm{x})^{\top}\bm{B}(\bm{x})^{-1}\bm{J}^{\mathrm{ss}}(\bm{x})}{P^{\mathrm{ss}}(\bm{x})}.\label{eq:I0_def}
\end{align}
By using Eqs.~\eqref{eq:TCRI} and \eqref{eq:I0_def}, the thermodynamic uncertainty relation is
obtained as follows:
\begin{equation}
\frac{\mathrm{Var}_{\theta=0}\left[\Theta_{\mathrm{cur}}(\bm{\Gamma})\right]}{\jmath^{2}}\ge\frac{2}{\Delta S_{\mathrm{tot}}},\label{eq:TUR_ineq}
\end{equation}
where $\Delta S_{\mathrm{tot}}$ is the total entropy production:
\begin{equation}
\Delta S_{\mathrm{tot}}\equiv T\int d\bm{x}\,\frac{\bm{J}^{\mathrm{ss}}(\bm{x})^{\top}\bm{B}(\bm{x})^{-1}\bm{J}^{\mathrm{ss}}(\bm{x})}{P^{\mathrm{ss}}(\bm{x})}.\label{eq:Stot_def}
\end{equation}
Equation~\eqref{eq:Stot_def} is the total entropy production assuming that all variables are
even under time reversal. 
When systems include odd variables (e.g., underdamped systems), 
the total entropy production is expressed differently.
In particular, the thermodynamic uncertainty relation including only the total entropy production term
 is violated in underdamped systems \cite{Fischer:2018:UTUR, Vu:2019:UTUR}.
Calculations above shows that, from the perspective of statistical inference, the current
$\Theta_{\mathrm{cur}}(\bm{\Gamma})$ is an estimator which infers
$\theta$ and the total entropy production corresponds to the Fisher
information in $\theta$-space. The Fisher information describes the
log-likelihood change when varying a parameter $\theta$. If the change
is large, the curvature of the log-likelihood becomes steeper, which
results in a more accurate parameter inference.

\subsection{Equality condition near equilibrium}

Identifying the thermodynamic uncertainty relation as the Cram\'er--Rao inequality, we can obtain
the equality condition of the thermodynamic uncertainty relation, which was not reported in the approach
based on the fluctuation-response inequality. From the equality condition of the Cram\'er--Rao
inequality, Eq.~\eqref{eq:TCRI} is satisfied with equality if
and only if the following relation holds
\begin{equation}
\frac{\partial}{\partial\theta}\ln\mathcal{P}_{\theta}(\bm{\Gamma})=\mu(\theta)\left[\Theta(\bm{\Gamma})-\psi(\theta)\right],\label{eq:CRI_eq_cond}
\end{equation}
where $\mu(\theta)$ is a scaling function (Eq.~\eqref{eq:equality_condition}
in Appendix~\ref{sec:info_ineqs}). When this relation holds, 
the thermodynamic uncertainty relation also holds with equality. For simplicity, we first consider
a one-dimensional system with periodic boundary conditions. 
Converting
from Ito to Stratonovich-type currents {[}cf. Eq.~\eqref{eq:JS_JI_result}
in Appendix~\ref{sec:currents}{]}, the left-hand side of Eq.~\eqref{eq:CRI_eq_cond} is
\begin{align}
 & \frac{\partial}{\partial\theta}\ln\mathcal{P}_{\theta}(\Gamma)\nonumber \\
 & =\frac{1}{2}\int_{0}^{T}dt\,\left[\frac{J^{\mathrm{ss}}}{P^{\mathrm{ss}}(x)B(x)}\bullet\dot{x}-\frac{J^{\mathrm{ss}}A_{\theta}(x)}{P^{\mathrm{ss}}(x)B(x)}\right]\nonumber \\
 & =\frac{1}{2}\int_{0}^{T}dt\,\frac{J^{\mathrm{ss}}}{P^{\mathrm{ss}}(x)B(x)}\circ\dot{x}-\frac{1+\theta}{2}\int_{0}^{T}dt\,\frac{[J^{\mathrm{ss}}]^{2}}{P^{\mathrm{ss}}(x)^{2}B(x)}.\label{eq:log_ll_last_line}
\end{align}
Accordingly, the right-hand side of the same equation becomes 
\begin{align}
    & \mu(\theta)\left[\Theta_{\mathrm{cur}}(\Gamma)-\psi(\theta)\right]\nonumber\\
    & =\mu(\theta)\left[\int_{0}^{T}dt\,\Lambda(x)\circ\dot{x}-(1+\theta)\jmath\right],\label{eq:log_ll_2}
\end{align}
where we used Eq.~\eqref{eq:Theta_cur_expec_def} in the last line
($\psi(\theta)=\left\langle \Theta_{\mathrm{cur}}(\Gamma)\right\rangle _{\theta}=(1+\theta)\jmath$).
Without loss of generality, we set $\mu(\theta)=J^{\mathrm{ss}}/2$
because multiplying $\Lambda(x)$ with a constant results in the same
bound. Correspondence between Eqs.~\eqref{eq:log_ll_last_line} and
\eqref{eq:log_ll_2} should hold for an arbitrary trajectory $\Gamma$
to attain  equality in the thermodynamic uncertainty relation. 
From Eqs.~\eqref{eq:CRI_eq_cond}--\eqref{eq:log_ll_2}, we determine that for an arbitrary $\Gamma$,
 the following
relation should be satisfied: 
\begin{equation}
\Xi(\Gamma)-(1+\theta)J^{\mathrm{ss}}\Psi(\Gamma)=0,\label{eq:eqcondition2_1}
\end{equation}
where 
\begin{align}
\Xi(\Gamma) & \equiv\int_{0}^{T}dt\,\left[\Lambda(x)-\frac{1}{P^{\mathrm{ss}}(x)B(x)}\right]\circ\dot{x},\label{eq:Xi_def}\\
\Psi(\Gamma) & \equiv T\int dx\Lambda(x)-\int_{0}^{T}\frac{1}{P^{\mathrm{ss}}(x)^{2}B(x)}dt.\label{eq:Psi_def}
\end{align}
Equation~\eqref{eq:eqcondition2_1} is a necessary and sufficient condition 
for the thermodynamic uncertainty relation to hold with equality. 
To satisfy Eq.~\eqref{eq:eqcondition2_1}, 
$\Xi(\Gamma)$ and $(1+\theta)J^\mathrm{ss}\Psi(\Gamma)$ should individually vanish \footnote{
Terms with and without $\dot{x}$ cannot cancel each other. 
Suppose cancelling both terms is possible, i.e., there exist functions $h_1(x)$ and $h_2(x)$ such that $\int_0^T h_1(x)\circ \dot{x} dt = \int_0^T h_2(x) dt$ for all $\Gamma$ (we exclude a trivial case $h_1(x)=h_2(x)=0$). 
Such a condition yields $h_1(x)\circ \dot{x} - h_2(x)=0$, indicating that we can determine $\dot{x}$
deterministically given $x$. However, owing to the stochastic nature of Langevin equations, such a condition is
not possible. }. 
From the condition $\Xi(\Gamma) = 0$, 
we find that the only quantity that satisfies the equality is the current:
\begin{equation}
\Theta_{\mathrm{tot}}(\Gamma)\equiv\int_{0}^{T}dt\,\frac{1}{P^{\mathrm{ss}}(x)B(x)}\circ\dot{x}.\label{eq:Theta_tot_def}
\end{equation}
For $B(x)=D$ (additive noise), $\Theta_{\mathrm{tot}}(\Gamma)$ is
$\Theta_{\mathrm{tot}}(\Gamma)\propto\int_{0}^{T}dt\,\dot{s}_{\mathrm{tot}}(t),$
where $\dot{s}_{\mathrm{tot}}$ is the stochastic total entropy production
rate: 
\begin{align}
\dot{s}_{\mathrm{tot}} & \equiv\frac{\dot{q}}{D}-\frac{d}{dt}\ln P^{\mathrm{ss}}(x)\nonumber\\
 & =\frac{A(x)\circ\dot{x}}{D}-\frac{\partial_{x}P^{\mathrm{ss}}(x)}{P^{\mathrm{ss}}(x)}\circ\dot{x}=\frac{J^{\mathrm{ss}}}{P^{\mathrm{ss}}(x)D}\circ\dot{x}.\label{eq:stochastic_stot_def}
\end{align}
Here, $\dot{q}\equiv A(x)\circ\dot{x}$ is the stochastic heat dissipation
rate \cite{Seifert:2012:FTReview}. Although the first term $\Xi(\Gamma)$
in Eq.~\eqref{eq:eqcondition2_1} vanishes with $\Theta_{\mathrm{tot}}(\Gamma)$, 
the second term
does not. Still, when we consider a near-equilibrium condition, $J^{\mathrm{ss}}\to0$,
the second term in Eq.~\eqref{eq:eqcondition2_1} converges to $0$.
Our result shows that $\Theta_{\mathrm{tot}}(\Gamma)$ (and its
multiples) is the only quantity which can attain equality
near equilibrium. This also holds for multi-dimensional systems with
$\bm{B}(\bm{x})=\bm{D}$ (additive noise). Particularly, converting
from Ito to Stratonovich-type current {[}Eq.~\eqref{eq:JS_JI_mult_result}
in Appendix~\ref{sec:currents}{]}, we repeat the same calculations as for the one-dimensional case
to obtain 
\begin{align}
\ln\mathcal{P}_{\theta}(\bm{\Gamma}) & =\frac{1}{2}\int_{0}^{T}dt\frac{\bm{J}^{\mathrm{ss}}(\bm{x})^{\top}\bm{D}^{-1}}{P^{\mathrm{ss}}(\bm{x})}\circ\dot{\bm{x}}\nonumber \\
 & -\frac{1+\theta}{2}\int_{0}^{T}dt\,\frac{\bm{J}^{\mathrm{ss}}(\bm{x})^{\top}\bm{D}^{-1}\bm{J}^{\mathrm{ss}}(\bm{x})}{P^{\mathrm{ss}}(\bm{x})^{2}}.\label{eq:mult_lnP}
\end{align}
Equation~\eqref{eq:mult_lnP} is the multi-dimensional generalization
of Eq.~\eqref{eq:log_ll_last_line}. Again, we define the following
current: 
\begin{equation}
\Theta_{\mathrm{tot}}(\bm{\Gamma})\equiv\int_{0}^{T}\frac{\bm{J}^{\mathrm{ss}}(\bm{x})^{\top}\bm{D}^{-1}}{P^{\mathrm{ss}}(\bm{x})}\circ\dot{\bm{x}}dt.\label{eq:Theta_tot_mult_def}
\end{equation}
Repeating the analysis of the one-dimensional case, we find that the
equality of the thermodynamic uncertainty relation is satisfied near equilibrium if and only if the
current is Eq.~\eqref{eq:Theta_tot_mult_def} (and its multiples).
$\Theta_{\mathrm{tot}}(\bm{\Gamma})$ of Eq.~\eqref{eq:Theta_tot_mult_def}
satisfies $\Theta_{\mathrm{tot}}(\bm{\Gamma})=\int_{0}^{T}\dot{s}_{\mathrm{tot}}dt$,
where $\dot{s}_{\mathrm{tot}}$ is the stochastic total entropy production
rate for multi-dimensional systems: 
\begin{equation}
\dot{s}_{\mathrm{tot}}\equiv\bm{A}(\bm{x})^{\top}\bm{D}^{-1}\circ\dot{\bm{x}}-\frac{d}{dt}\ln P^{\mathrm{ss}}(\bm{x})=\frac{\bm{J}^{\mathrm{ss}}(\bm{x})^{\top}\bm{D}^{-1}}{P^{\mathrm{ss}}(\bm{x})}\circ\dot{\bm{x}}.\label{eq:Stot_mult_def}
\end{equation}
In Eq.~\eqref{eq:Stot_mult_def}, $\bm{A}(\bm{x})^{\top}\bm{D}^{-1}\circ\dot{\bm{x}}$
corresponds to the stochastic medium entropy rate. The stochastic
total entropy production has been shown to attain equality near
equilibrium in Ref.~\cite{Pigolotti:2017:EP}, but it was not shown
that it is the only quantity which attains equality. This current was shown
to satisfy the equality of the thermodynamic uncertainty relation using the linear response \cite{Macieszczak:2018:TURLR},
and was demonstrated to provide the tightest quadratic bound for the
rate function \cite{Gingrich:2016:TUP}.

\subsection{Exact equality condition}

As the equality condition discussed above is asymptotic with respect
to $J^{\mathrm{ss}}\to0$, the equality is not met exactly. Next,
we seek an exact equality condition for the one-dimensional case.
With $\Theta_{\mathrm{tot}}(\Gamma)$ {[}Eq.~\eqref{eq:Theta_tot_def}{]},
the first term $\Xi(\Gamma)$ in Eq.~\eqref{eq:eqcondition2_1} vanishes,
while the second term $(1+\theta)J^{\mathrm{ss}}\Psi(\Gamma)$ does
not. Therefore, $\Psi(\Gamma)=0$ should hold for an arbitrary $\Gamma$
to attain equality in the thermodynamic uncertainty relation. In Eq.~\eqref{eq:Psi_def}, the
first term is a constant due to the integration with respect to $x$,
but the second term $\int_{0}^{T}1/[P^{\mathrm{ss}}(x)^{2}B(x)]dt$
depends on $\Gamma$. Therefore, to satisfy $\Psi(\Gamma)=0$ for
an arbitrary $\Gamma$, the integrand $1/[P^{\mathrm{ss}}(x)^{2}B(x)]$
should be constant, which yields 
\begin{equation}
P^{\mathrm{ss}}(x)=\frac{c}{\sqrt{B(x)}},\label{eq:Pss_eq_cond}
\end{equation}
with $c>0$ being a normalization constant. Indeed, substituting Eq.~\eqref{eq:Pss_eq_cond}
into $\Psi(\Gamma)$ {[}Eq.~\eqref{eq:Psi_def}{]}, we find $\Psi(\Gamma)=0$.
From $J^{\mathrm{ss}}=A(x)P^{\mathrm{ss}}(x)-\partial_{x}B(x)P^{\mathrm{ss}}(x)$,
we obtain $A(x)$ as follows: 
\begin{equation}
A(x)=\frac{J^{\mathrm{ss}}+\partial_{x}B(x)P^{\mathrm{ss}}(x)}{P^{\mathrm{ss}}(x)}=\kappa C(x)+C(x)\frac{d}{dx}C(x),\label{eq:Ax_condition}
\end{equation}
where $\kappa$ is an arbitrary parameter. $\Lambda(x)$ is given
by 
\begin{align}
\Lambda(x) & \propto\frac{1}{P^{\mathrm{ss}}(x)B(x)}\propto\frac{1}{\sqrt{B(x)}}=\frac{1}{C(x)}.\label{eq:Lambda_cond_supp}
\end{align}
Regardless of the system being near equilibrium, equality in the
thermodynamic uncertainty relation is attained if and only if Eqs.~\eqref{eq:Ax_condition} and
\eqref{eq:Lambda_cond_supp} hold, which has not been demonstrated in the litarature.

For the multi-dimensional case with $\bm{B}(\bm{x})=\bm{D}$, from
Eq.~\eqref{eq:mult_lnP}, the exact equality of the thermodynamic uncertainty relation is satisfied
if and only if the current is proportional to $\Theta_{\mathrm{tot}}(\bm{\Gamma})$
{[}Eq.~\eqref{eq:Theta_tot_mult_def}{]} and $\bm{J}^{\mathrm{ss}}(\bm{x})\bm{D}^{-1}\bm{J}^{\mathrm{ss}}(\bm{x})/P^{\mathrm{ss}}(\bm{x})^{2}$
is constant for any $\bm{x}$. However, it is difficult to specify
systems satisfying the latter condition. 

\subsection{Example: Particle in a periodic potential}

We numerically confirm these equality conditions by considering a
particle on a periodic potential with a period of $2\pi$ subject to 
an external force. We consider the following periodic drift:
\begin{equation}
A(x)=(a+\sin(x))(b+\cos(x)),\label{eq:Ax_periodic_def}
\end{equation}
where $a>1$ and $b\ge0$ are model parameters. 
The drift defined by Eq.~\eqref{eq:Ax_periodic_def} is the sum of the periodic potential and the external force. 
Let $V(x)$ be an effective potential function
of Eq.~\eqref{eq:Ax_periodic_def}:
\begin{align}
V(x) & \equiv-\int A(x)dx\nonumber \\
 & =-\frac{1}{2}(a+\sin(x))^{2}-b(ax-\cos(x)).\label{eq:potential_V_def}
\end{align}
In Fig.~\ref{fig:FIG1}(b), $V(x)$ is plotted for $b=0$ (solid
line), $0.5$ (dashed line), and $1.0$ (dot-dashed line) with $a=2$.
When $b=0$, because $V(x)$ does not have a tilt,
the system
is in equilibrium at a steady state. 
We employ the drift of Eq.~\eqref{eq:Ax_periodic_def} since $C(x)$
satisfying Eq.~\eqref{eq:Ax_condition} can be expressed analytically.
From Eq.~\eqref{eq:TUR_ineq}, we define 
\begin{equation}
\mathcal{F}\equiv\frac{\mathrm{Var}_{\theta=0}[\Theta(\Gamma)]\Delta S_{\mathrm{tot}}}{\jmath^{2}}\ge2,\label{eq:mathcal_F_def}
\end{equation}
which is $\mathcal{F}=2$ for the equality of the thermodynamic uncertainty relation. We numerically
calculate $\mathcal{F}$ by repeating simulations $N_{S}=5.0\times10^{6}$
times with temporal resolution $\Delta t=0.0002$ (parameters are
shown in the caption of Fig.~\ref{fig:FIG1}(c)). We consider the
following four cases: (i) $\Lambda(x)=1/[P^{\mathrm{ss}}(x)B(x)]$,
$C(x)=\sqrt{D}$, (ii) $\Lambda(x)=1/C(x)\propto1/[P^{\mathrm{ss}}(x)B(x)]$,
$C(x)=a+\sin(x)$, (iii) $\Lambda(x)=1$, $C(x)=\sqrt{D}$, and (iv)
$\Lambda(x)=A(x)$, $C(x)=\sqrt{D}$ (see Appendix~\ref{sec:periodic_potential}).
Cases (i) and (ii) satisfy the near-equilibrium equality condition
and (ii) further satisfies Eq.~\eqref{eq:Ax_condition}, while (iii)
and (iv) do not. The current of (iv) depicts the stochastic heat dissipation.
Figure~\ref{fig:FIG1}(c) shows $\mathcal{F}$ as a function of $J^{\mathrm{ss}}$
for (i), (ii), (iii), and (iv), which are described by circles, diamonds,
triangles, and squares, respectively. For $J^{\mathrm{ss}}\to0$,
only (i) and (ii) show $\mathcal{F}\to2$. Case (ii) exhibits $\mathcal{F}\simeq2$
for all $J^{\mathrm{ss}}$, indicating that it satisfies the equality
even when the system is far from equilibrium, as expected. Cases (iii)
and (iv) are $\mathcal{F}>2$; thus, they do not satisfy the equality
of the thermodynamic uncertainty relation, which agrees with our theoretical result.

\section{Chapman--Robbins inequality}

\subsection{Derivation of uncertainty relation}

A different information inequality can be applied to the system from
the identification of thermodynamic systems in terms of statistical
inference. By applying the Chapman--Robbins inequality \cite{Kendall:ClassicalInference:2A,George:2001:StatInfer,Lehmann:2003:EstimationBook}
(Appendix~\ref{sec:info_ineqs}), which is a generalization of the
Cram\'er--Rao inequality, to Eq.~\eqref{eq:dynamics_def}, the following relation holds:
\begin{equation}
\frac{\mathrm{Var}_{\theta=0}\left[\Theta(\bm{\Gamma})\right]}{\left[\left\langle \Theta(\bm{\Gamma})\right\rangle _{\theta}-\left\langle \Theta(\bm{\Gamma})\right\rangle _{\theta=0}\right]^{2}}\ge\frac{1}{\mathbb{D}_{\mathrm{PE}}\left[\mathcal{P}_{\theta}||\mathcal{P}_{\theta=0}\right]},\label{eq:HCRI}
\end{equation}
where $\mathbb{D}_{\mathrm{PE}}\left[\mathcal{P}_{\theta}||\mathcal{P}_{\theta=0}\right]$
is the Pearson divergence between $\mathcal{P}_{\theta}$ and $\mathcal{P}_{\theta=0}$
defined by 
\begin{equation}
\mathbb{D}_{\mathrm{PE}}\left[\mathcal{P}_{\theta}||\mathcal{P}_{\theta=0}\right]\equiv\int\mathcal{D}\bm{\Gamma}\,\left(\frac{\mathcal{P}_{\theta}(\bm{\Gamma})}{\mathcal{P}_{\theta=0}(\bm{\Gamma})}-1\right)^{2}\mathcal{P}_{\theta=0}(\bm{\Gamma}).\label{eq:Pearson_div_def}
\end{equation}
In Eq.~\eqref{eq:Pearson_div_def}, $\left[\left\langle \Theta(\bm{\Gamma})\right\rangle _{\theta}-\left\langle \Theta(\bm{\Gamma})\right\rangle _{\theta=0}\right]^{2}$
describes the difference between two dynamics characterized by $\theta=0$
and $\theta\ne0$, which represents the sensitivity of the system.
Thus, the ratio between the variance of the unperturbed dynamics and the
sensitivity is bounded from below by the reciprocal of the Pearson
divergence between the two dynamics. For $\theta\to0$, Eq.~\eqref{eq:HCRI}
reduces to the Cram\'er--Rao inequality {[}Eq.~\eqref{eq:TCRI}{]}
and the fluctuation-response inequality. Although the fluctuation-response inequality only holds for sufficiently weak perturbations,
as it holds locally around $\theta=0$, Eq.~\eqref{eq:HCRI} is satisfied
for an arbitrary $\theta\ne0$, indicating that Eq.~\eqref{eq:HCRI}
can be used beyond a linear response regime. In stochastic thermodynamics,
by using thermodynamic inequalities, several measures of efficiency have been
calculated to evaluate the performance of systems \cite{Barato:2014:CellInfo,Hartich:2016:SensCap,Goldt:2017:STLearning}.
Similarly, we can evaluate the efficiency in terms of sensitivity
and precision with Eq.~\eqref{eq:HCRI}. As Eq.~\eqref{eq:HCRI}
holds for an arbitrary $\theta$, the Chapman--Robbins inequality
is often stated to provide the lower bound, which is at least as tight
as the Cram\'er--Rao inequality: 
\begin{align}
    \mathrm{Var}_{\theta=0}\left[\Theta(\bm{\Gamma})\right]&\ge\sup_{\theta}\frac{\left[\left\langle \Theta(\bm{\Gamma})\right\rangle _{\theta}-\left\langle \Theta(\bm{\Gamma})\right\rangle _{\theta=0}\right]^{2}}{\mathbb{D}_{\mathrm{PE}}\left[\mathcal{P}_{\theta}||\mathcal{P}_{\theta=0}\right]}\nonumber\\
    &\ge\frac{[{\displaystyle \partial_{\theta}\left\langle \Theta(\bm{\Gamma})\right\rangle _{\theta}}]_{\theta=0}^{2}}{\mathcal{I}(0)}.\label{eq:HCRI2}
\end{align}
However, in the present manuscript, we only focus on the relation of
Eq.~\eqref{eq:HCRI}.

\subsection{Example 1: Linear Langevin equation}

First, we study Eq.~\eqref{eq:HCRI} in a linear Langevin equation
because the Pearson divergence can be obtained analytically. We consider
the following equation in Eq.~\eqref{eq:dynamics_def}: 
\begin{equation}
A_{\theta}(x,t)=-\alpha x+\theta u(t),\hspace*{1em}C(x,t)=\sqrt{D},\label{eq:A_linear_def}
\end{equation}
where $\alpha>0$, $u(t)$ is an arbitrary input function, and $D$
is the noise intensity. The initial condition is $x=0$ at $t=0$.
By following the calculation of the path integral, the Pearson divergence
is represented analytically by (Appendix~\ref{sec:OU_bound}) 
\begin{equation}
\mathbb{D}_{\mathrm{PE}}\left[\mathcal{P}_{\theta}||\mathcal{P}_{\theta=0}\right]=-1+\exp\left(\frac{\theta^{2}}{2D}\int_{0}^{T}u(t)^{2}dt\right).\label{eq:DPE_OU}
\end{equation}
When we define $\Theta_{x}(\Gamma)\equiv\int_{0}^{T}\dot{x}dt$, $\Theta_{x}(\Gamma)$
is the position $x(T)$. Therefore, the Chapman--Robbins
inequality in Eq.~\eqref{eq:HCRI} is 
\begin{equation}
\frac{\mathrm{Var}_{\theta=0}\left[x(T)\right]}{\left[\left\langle x(T)\right\rangle _{\theta}-\left\langle x(T)\right\rangle _{\theta=0}\right]^{2}}\ge\frac{1}{\mathbb{D}_{\mathrm{PE}}\left[\mathcal{P}_{\theta}||\mathcal{P}_{\theta=0}\right]}.\label{eq:linear_ChapmanRobbins}
\end{equation}
We also consider the lower bound of the fluctuation-response inequality {[}Eq.~\eqref{eq:FRI_def}{]}:
$1/\left[\theta^{2}\mathcal{I}(0)\right]=2D/\left[\theta^{2}\int_{0}^{T}u(t)^{2}dt\right]$.
Using $\exp(x)\ge1+x$ in Eq.~\eqref{eq:DPE_OU}, we can easily show
\begin{equation}
\mathbb{D}_{\mathrm{PE}}\left[\mathcal{P}_{\theta}||\mathcal{P}_{\theta=0}\right]\ge\frac{\theta^{2}}{2D}\int_{0}^{T}u(t)^{2}dt=\theta^{2}\mathcal{I}(0).\label{eq:DPE_ineq_1}
\end{equation}
When $x=0$ at time $t=0$, the variance and the mean of Eq.~\eqref{eq:A_linear_def}
are given by 
\begin{align}
\left\langle x(T)\right\rangle _{\theta} & =\left[\theta\int_{0}^{T}u(t)e^{\alpha t}dt\right]e^{-\alpha T},\label{eq:linear_xT_expectation}\\
\mathrm{Var}_{\theta=0}[x(T)] & =\frac{D}{\alpha}\left[1-e^{-2\alpha T}\right].\label{eq:linear_xT_Var}
\end{align}
For any $u(t)\ge0$, using the Cauchy--Schwarz inequality, we have
\begin{align}
\left[\int_{0}^{T}u(t)e^{\alpha t}dt\right]^{2} & \le\int_{0}^{T}u(t)^{2}dt\int_{0}^{T}e^{2\alpha t}dt\nonumber \\
 & =\frac{e^{2\alpha T}-1}{2\alpha}\int_{0}^{T}u(t)^{2}dt,\label{eq:linear_CauchySchwarz}
\end{align}
which yields the following relation: 
\begin{equation}
\frac{\mathrm{Var}_{\theta=0}\left[x(T)\right]}{\left[\left\langle x(T)\right\rangle _{\theta}-\left\langle x(T)\right\rangle _{\theta=0}\right]^{2}}\ge\frac{1}{\theta^{2}\mathcal{I}(0)}\ge\frac{1}{\mathbb{D}_{\mathrm{PE}}\left[\mathcal{P}_{\theta}||\mathcal{P}_{\theta=0}\right]}.\label{eq:OU_ineq_relation}
\end{equation}
 Equation~\eqref{eq:OU_ineq_relation} indicates that the bound
of the fluctuation-response inequality is always tighter than that of the Chapman--Robbins inequality.
This relation seems to be inconsistent with the fact that the Chapman--Robbins
inequality is at least as tight as the Cram\'er--Rao inequality
{[}cf. Eq.~\eqref{eq:HCRI2}{]}. However, Eq.~\eqref{eq:HCRI2}
does not contradict Eq.~\eqref{eq:OU_ineq_relation}. Although
the fluctuation-response inequality always holds for the linear system, it is violated in nonlinear
systems, as shown in the next example.

\subsection{Example 2: Limit cycle oscillator}

Equation~\eqref{eq:HCRI} bounds the sensitivity and the precision,
which is of particular interest in limit cycle oscillators. Circadian
clocks are biological limit cycle oscillators in organisms which orchestrate
the activities of several organs. Their temporal precision is incredibly
high (the standard deviation of the period is $3$--$5$ min over $24$~h)
\cite{Moortgat:2000:Precision} and several mechanisms have been
proposed for such precision \cite{Winfree:2001:GeoBiolTime,Needleman:2001:CEP,Kori:2012:OscReg,Hasegawa:2018:CEP}.
Simultaneously, circadian clocks have to synchronize to sunlight cycles
such that biological activities are operational at specific times.
As oscillators with higher sensitivity are vulnerable to periodic
signals as well as noise, precision and sensitivity appear to be trade-off
factors, which is an uncertainty relation in stochastic oscillators
\cite{Hasegawa:2013:OptimalPRC,Hasegawa:2014:PRL,Fei:2018:Osc}.

\begin{figure}
\includegraphics[width=8.5cm]{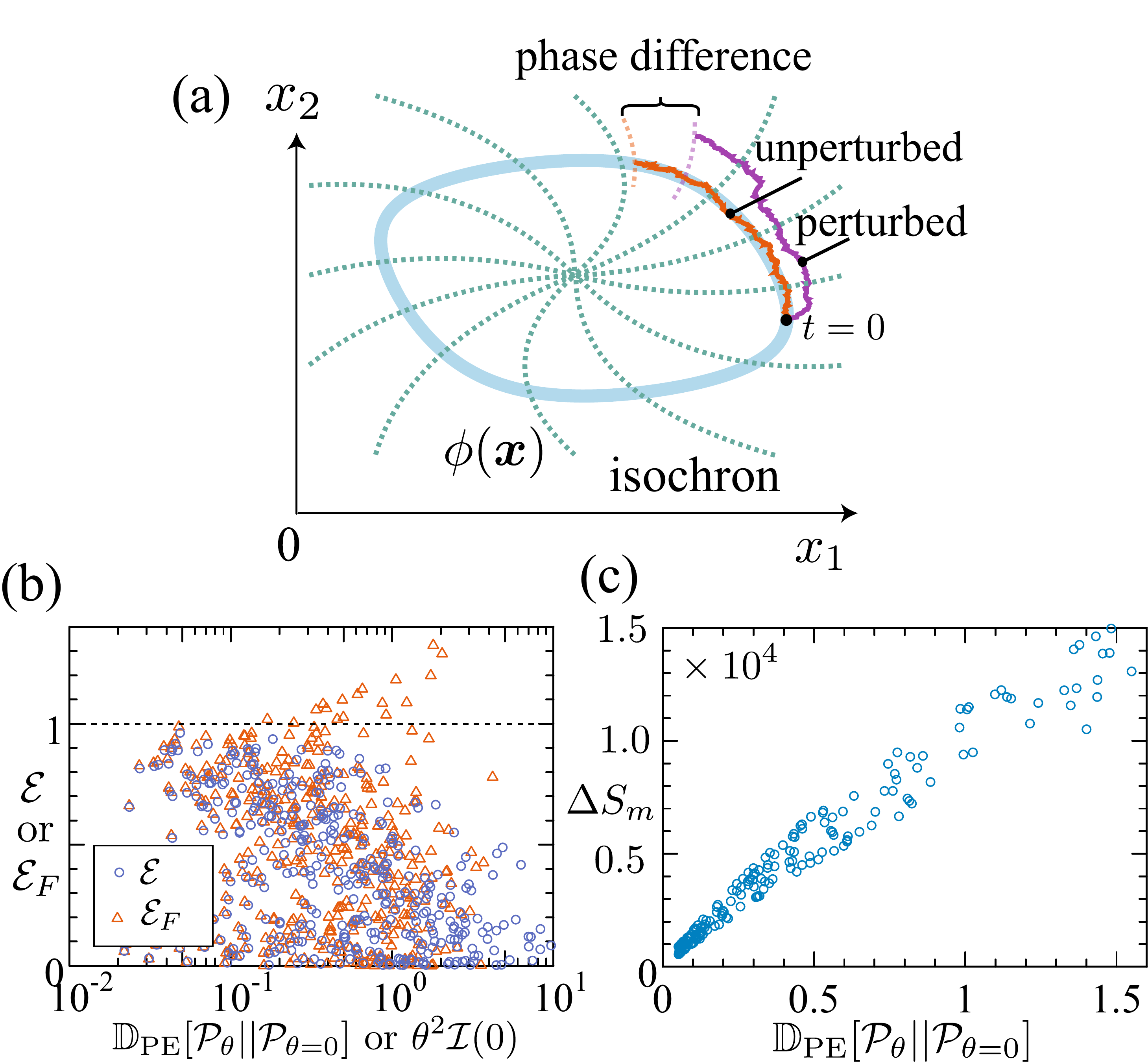}

\caption{(a) The phase $\phi(\bm{x})$ is defined on the coordinate space and the
isochron (green dotted lines) denotes the equiphase surface. The phase
sensitivity is quantified by the phase difference between perturbed
(purple line) and unperturbed (orange line) dynamics, with the deterministic
oscillation shown by the light blue solid line. (b) Numerical verification
of the Chapman--Robbins inequality in the stochastic oscillator.
For random realizations, $\mathcal{E}$ (blue circles) and $\mathcal{E}_{F}$
(orange triangles) are plotted as a function of $\mathbb{D}_{\mathrm{PE}}\left[\mathcal{P}_{\theta}||\mathcal{P}_{\theta=0}\right]$
and $\theta^{2}\mathcal{I}(0)$, respectively. When the Chapman--Robbins
inequality and the fluctuation-response inequality are satisfied, $\mathcal{E}\le1$ and $\mathcal{E}_{F}\le1$,
respectively. Parameters are $D\in[0.05,0.2]$, $\theta\in[0.1,0.5]$,
$T\in\{\tau/4,\tau/8,\tau/16\}$. (c) Medium entropy $\Delta S_{m}$
as a function of $\mathbb{D}_{\mathrm{PE}}\left[\mathcal{P}_{\theta}||\mathcal{P}_{\theta=0}\right]$,
where circles denote random realizations. Parameters are $D\in[0.01,0.2]$,
$\theta=0.05$, $T=\tau$ ($\tau$ is the period of the deterministic
oscillation). \label{fig:isochrone}}
\end{figure}
We consider a deterministic limit cycle oscillator defined by $\dot{\bm{x}}=\bm{A}(\bm{x})$.
We can define the phase $\phi$ on a closed orbit of the deterministic
oscillation by $\dot{\phi}=\Omega$, where $\Omega\equiv2\pi/\tau$
is the angular frequency of the oscillation ($\tau$ is the period
of the deterministic oscillation). In the presence of an external
signal, the dynamics obey Eq.~\eqref{eq:dynamics_def} with 
\begin{equation}
\bm{A}_{\theta}(\bm{x},t)=\bm{A}(\bm{x})+\theta\bm{u}(t),
\label{eq:limitcycle_A_def}
\end{equation}
where $\bm{u}(t)=[u_{1}(t),...,u_{N}(t)]^{\top}$ depicts the signal.
Although $\phi$ is defined only on the deterministic closed orbit,
we can expand the definition of the phase over the entire $\bm{x}$
space, which is denoted by $\phi(\bm{x})$ \cite{Kuramoto:2003:OscBook}.
$\phi(\bm{x})$ can be calculated by directly solving the ordinary
differential equation (Appendix~\ref{sec:simulations}). The integrated
phase from $t=0$ to $t=T$ is given by $\int_{0}^{T}\dot{\phi}(\bm{x}(t))dt$.
Since the time derivative of $\phi$ is $\dot{\phi}=\sum_{i=1}^{N}\left(\partial_{x_{i}}\phi(\bm{x})\right)\circ\dot{x}_{i}$,
we define a current 
\begin{equation}
\Theta_{\phi}(\bm{\Gamma})\equiv\int_{0}^{T}\sum_{i=1}^{N}\left(\frac{\partial}{\partial x_{i}}\phi(\bm{x})\right)\circ\dot{x}_{i}dt,
\label{eq:limitcycle_Theta_phi_def}
\end{equation}
which is the integrated phase calculated from a trajectory $\bm{\Gamma}$.
The temporal precision of the oscillator is quantified by $\mathrm{Var}_{\theta=0}\left[\Theta_{\phi}(\bm{\Gamma})\right]$,
which is the variance of the phase of unperturbed dynamics 
(lower variance corresponds to higher precision). The sensitivity
of the oscillator can be quantified by the phase difference between
perturbed and unperturbed dynamics. Therefore, we define the phase
sensitivity as $\left[\left\langle \Theta_{\phi}(\bm{\Gamma})\right\rangle _{\theta}-\left\langle \Theta_{\phi}(\bm{\Gamma})\right\rangle _{\theta=0}\right]^{2}$.
Figure~\ref{fig:isochrone}(a) illustrates the phase $\phi(\bm{x})$
(the dotted line shows the isochron) and the phase difference between
unperturbed and perturbed dynamics. From Eq.~\eqref{eq:HCRI}, 
the phase variance and the sensitivity
satisfy the relation: 
\begin{equation}
\frac{\mathrm{Var}_{\theta=0}\left[\Theta_{\phi}(\bm{\Gamma})\right]}{\left[\left\langle \Theta_{\phi}(\bm{\Gamma})\right\rangle _{\theta}-\left\langle \Theta_{\phi}(\bm{\Gamma})\right\rangle _{\theta=0}\right]^{2}}\ge\frac{1}{\mathbb{D}_{\mathrm{PE}}\left[\mathcal{P}_{\theta}||\mathcal{P}_{\theta=0}\right]},\label{eq:Pphi_Sphi_ineq}
\end{equation}
which shows that as $\mathbb{D}_{\mathrm{PE}}\left[\mathcal{P}_{\theta}||\mathcal{P}_{\theta=0}\right]$
increases, both precision and sensitivity are improved simultaneously.
In limit cycle oscillators, perturbations are often applied to experimentally
observe the properties of oscillators. Particularly, the sensitivity
$\left[\left\langle \Theta_{\phi}(\bm{\Gamma})\right\rangle _{\theta}-\left\langle \Theta_{\phi}(\bm{\Gamma})\right\rangle _{\theta=0}\right]^{2}$ corresponds to the square of the phase response
curve \cite{Schwemmer:2011:PRCBook}. The phase variance 
and the sensitivity are common measures and
were used in Refs.~\cite{Hasegawa:2013:OptimalPRC,Hasegawa:2014:PRL,Fei:2018:Osc} to study their trade-off
relation. However, an explicit inequality between the two quantities
has not been reported yet. We determine that their ratio is lower
bounded by the Pearson divergence, which is an information quantity.
Such information quantities play important roles in information
thermodynamics \cite{Parrondo:2015:InfoThermo}. The Pearson divergence
between the original and perturbed trajectories is experimentally measurable
because trajectory-based quantities were previously measured \cite{Trepagnier:2004:Exp,Collin:2005:CrooksVerify,Schuler:2005:TwoLevelSystem,Tietz:2006:EPMeasure,Andrieux:2007:EPExp}.

We numerically confirm the inequality relation of Eq.~\eqref{eq:Pphi_Sphi_ineq}. We employ the Van der Pol oscillator \cite{VanDerPol:1926:VDP},
which is representative of many limit cycle oscillators, including
circadian oscillators. This oscillator has been extensively employed in the literature.
The noisy Van der Pol oscillator is defined by 
\begin{equation}
\bm{A}_{\theta}(\bm{x},t)=\left[\begin{array}{c}
x_{2}+\theta u(t)\\
\zeta(1-x_{1}^{2})x_{2}-x_{1}
\end{array}\right],\bm{B}=\left[\begin{array}{cc}
D & 0\\
0 & D
\end{array}\right],\label{eq:LCT_def}
\end{equation}
where $\zeta$ is a model parameter ($\zeta=2.5$ throughout), $D$
is noise intensity, and $u(t)=1$ for $t>0$ and $u(t)=0$ for $t\le0$.
Using Monte Carlo simulations, we solve the Langevin equation Eq.~\eqref{eq:LCT_def}
with time resolution $\Delta t=0.0002$ and evaluate Pearson divergence,
the sensitivity, and the phase variance (Appendix~\ref{sec:simulations}).
We randomly select $D$, $\theta$, and $T$, and repeat simulations
$N_{S}=5.0\times10^{5}$ times at each of the selected parameter settings
(ranges of the random parameters are shown in the caption of Fig.~\ref{fig:isochrone}(b)).
For initial values, we randomly select a point on the closed orbit
of the deterministic oscillation (the light blue line in Fig.~\ref{fig:isochrone}(a)).
We calculate 
\begin{equation}
\mathcal{E}\equiv\frac{\left[\left\langle \Theta_{\phi}(\bm{\Gamma})\right\rangle _{\theta}-\left\langle \Theta_{\phi}(\bm{\Gamma})\right\rangle _{\theta=0}\right]^{2}}{\mathrm{Var}_{\theta=0}\left[\Theta_{\phi}(\bm{\Gamma})\right]\mathbb{D}_{\mathrm{PE}}\left[\mathcal{P}_{\theta}||\mathcal{P}_{\theta=0}\right]},\label{eq:limitcycle_efficiency_def}
\end{equation}
which should be $\mathcal{E}\le1$ according to Eq.~\eqref{eq:Pphi_Sphi_ineq}.
Identifying $\mathbb{D}_{\mathrm{PE}}\left[\mathcal{P}_{\theta}||\mathcal{P}_{\theta=0}\right]$
as the cost, we can regard $\mathcal{E}$ as the efficiency of oscillators
(larger $\mathcal{E}$ corresponds to higher efficiency). In Fig.~\ref{fig:isochrone}(b),
we plot the random realizations of $\mathcal{E}$ (circles) as a function
of $\mathbb{D}_{\mathrm{PE}}\left[\mathcal{P}_{\theta}||\mathcal{P}_{\theta=0}\right]$.
For comparison, we define 
\begin{equation}
\mathcal{E}_{F}\equiv\frac{\left[\left\langle \Theta_{\phi}(\bm{\Gamma})\right\rangle _{\theta}-\left\langle \Theta_{\phi}(\bm{\Gamma})\right\rangle _{\theta=0}\right]^{2}}{\mathrm{Var}_{\theta=0}\left[\Theta_{\phi}(\bm{\Gamma})\right]\theta^{2}\mathcal{I}(0)},\label{eq:limitcycle_EF_def}
\end{equation}
which is based on the fluctuation-response inequality and plot $\mathcal{E}_{F}$ (triangles)
as a function of $\theta^{2}\mathcal{I}(0)$. When the fluctuation-response inequality is satisfied,
$\mathcal{E}_{F}\le1$. We observe that all circles are located below
$1$, indicating that the Chapman--Robbins inequality is satisfied
for all realizations. Conversely, some triangles are above 1, which
suggests violation of the fluctuation-response inequality. Although the linear response provides
an exact response for linear systems, it is accurate only
for sufficiently weak perturbations in the case of nonlinear systems.
Thus, the fluctuation-response inequality is violated for nonlinear cases.

When the stochastic oscillator is approximated linearly around the
deterministic orbit, Eq.~\eqref{eq:DPE_OU} shows that the Pearson
divergence increases exponentially as the noise intensity decreases.
It is known that entropy production increases when noise intensity
$D$ decreases \cite{Tome:2006:EPinFPE}. Therefore, lower noise
intensity increases both entropy production and Pearson divergence.
We numerically demonstrate a relation between $\mathbb{D}_{\mathrm{PE}}\left[\mathcal{P}_{\theta}||\mathcal{P}_{\theta=0}\right]$
and entropy production. Let $\Delta S_{m}$ be the medium entropy defined
by 
\begin{equation}
\Delta S_{m}\equiv\left\langle \frac{1}{D}\sum_{i=1}^{N}\int_{0}^{T}A_{i}(\bm{x})\circ\dot{x}_{i}dt\right\rangle .\label{eq:limitcycle_Sm_def}
\end{equation}
When $T$ is sufficiently large, the boundary term can be ignored
and $\Delta S_{m}\simeq\Delta S_{\mathrm{tot}}$. Following the foregoing
simulation procedure, we calculate $\Delta S_{m}$ and $\mathbb{D}_{\mathrm{PE}}\left[\mathcal{P}_{\theta}||\mathcal{P}_{\theta=0}\right]$
(parameter settings are shown in the caption for Fig.~\ref{fig:isochrone}(c)).
In Fig.~\ref{fig:isochrone}(c), we plot $\Delta S_{m}$ as a function
of $\mathbb{D}_{\mathrm{PE}}\left[\mathcal{P}_{\theta}||\mathcal{P}_{\theta=0}\right]$
for fixed $\theta$ and $T$. We observe that $\Delta S_{m}$ increases
when $\mathbb{D}_{\mathrm{PE}}\left[\mathcal{P}_{\theta}||\mathcal{P}_{\theta=0}\right]$
increases, showing that a larger Pearson divergence can be achieved
for larger entropy production. Using simulations, Ref.~\cite{Fei:2018:Osc}
showed that both higher precision and higher sensitivity are achieved
with higher entropy production, which is consistent with our results.

\subsection{Conclusion}

In this paper, we have applied information inequalities to systems
described by Langevin equations to obtain inequalities in stochastic
processes. We have identified that the thermodynamic uncertainty relation is a particular case of
the Cram\'er--Rao inequality. Furthermore, we have applied the Chapman--Robbins
inequality to the systems to show that the ratio between the variance
and the sensitivity is bounded from below by the Pearson divergence.
By bridging statistical inference theory and stochastic thermodynamic
systems, this study can provide a useful basis for further developments
with respect to thermodynamic bounds.

\section*{Acknowledgement}
This work was supported by MEXT KAKENHI
Grant No.~JP16K00325, JP19K12153. 

\appendix

\section{Information inequalities\label{sec:info_ineqs}}

Although information inequalities depict fundamental relations in
statistics and machine learning, they are less known in physics. Therefore,
we show their derivations for the readers' convenience \cite{Kendall:ClassicalInference:2A,George:2001:StatInfer,Lehmann:2003:EstimationBook}.

Let $X$ be a random variable, $P_{\theta}(X)$ be a probability density
function where $\theta$ is an arbitrary parameter, and $\Theta(X)$
be an unbiased estimator of $\psi(\theta)$ which indicates
$\psi(\theta) = \left\langle \Theta(X)\right\rangle _{\theta}$.
Then the following relation holds:
\begin{align}
 & \left\langle \left(\Theta(X)-\psi(\theta)\right)\left(\frac{\partial}{\partial\theta}\ln P_{\theta}(X)\right)\right\rangle _{\theta}\nonumber \\
 & =\int dX\left(\Theta(X)-\psi(\theta)\right)\left(\frac{\partial}{\partial\theta}\ln P_{\theta}(X)\right)P_{\theta}(X)\nonumber \\
 & =\frac{\partial}{\partial\theta}\left\langle \Theta(X)\right\rangle _{\theta}.\label{eq:CRI2}
\end{align}
Applying the Cauchy--Schwarz inequality to Eq.~\eqref{eq:CRI2},
we obtain the Cram\'er--Rao inequality: 
\begin{align}
\mathrm{Var}_{\theta}\left[\Theta(\Gamma)\right]&\ge\frac{\left[\partial_{\theta}\left\langle \Theta(X)\right\rangle _{\theta}\right]^{2}}{\left\langle \left(\partial_{\theta}\ln P_{\theta}(X)\right)^{2}\right\rangle _{\theta}}=\frac{\left[\partial_{\theta}\left\langle \Theta(X)\right\rangle _{\theta}\right]^{2}}{\left\langle -\partial_{\theta}^{2}\ln P_{\theta}(X)\right\rangle _{\theta}}\nonumber\\
&=\frac{\left[\partial_{\theta}\psi(\theta)\right]^{2}}{\mathcal{I}(\theta)},\label{eq:AppA_CRI}
\end{align}
where $\mathcal{I}(\theta)$ is the Fisher information. Its equality
condition is obtained from that of the Cauchy--Schwarz inequality.
From the left hand side of Eq.~\eqref{eq:CRI2}, if and only if the
following condition is satisfied,  the Cram\'er--Rao
inequality holds with equality: 
\begin{equation}
\frac{\partial}{\partial\theta}\ln P_{\theta}(X)=\mu\left[\Theta(X)-\psi(\theta)\right],\label{eq:equality_condition}
\end{equation}
where $\mu$ is a scaling parameter, which may depend on $\theta$
(i.e., $\mu=\mu(\theta)$).

The Chapman--Robbins inequality is a generalization of the Cram\'er--Rao
inequality. For $\vartheta\ne\theta$, we notice that 
\begin{align}
\left\langle \frac{P_{\vartheta}(X)}{P_{\theta}(X)}-1\right\rangle _{\theta} & =\int dX\,\left(P_{\vartheta}(X)-P_{\theta}(X)\right)=0.\label{eq:AppA_ChapRob_eq1}
\end{align}
Then the following relation holds: 
\begin{align}
 & \left\langle \left(\Theta(X)-\psi(\theta)\right)\left(\frac{P_{\vartheta}(X)}{P_{\theta}(X)}-1\right)\right\rangle _{\theta}\nonumber \\
 & =\int dX\left(\Theta(X)-\psi(\theta)\right)\left(\frac{P_{\vartheta}(X)}{P_{\theta}(X)}-1\right)P_{\theta}(X)\nonumber \\
 & =\left\langle \Theta(X)\right\rangle _{\vartheta}-\left\langle \Theta(X)\right\rangle _{\theta}.\label{eq:HCRI2_1}
\end{align}
Applying the Cauchy--Schwarz inequality to Eq.~\eqref{eq:HCRI2_1},
we obtain the Chapman--Robbins inequality: 
\begin{align}
\mathrm{Var}_{\theta}\left[\Theta(X)\right] & \ge\frac{\left[\left\langle \Theta(X)\right\rangle _{\vartheta}-\left\langle \Theta(X)\right\rangle _{\theta}\right]^{2}}{\left\langle \left(\frac{P_{\vartheta}(X)}{P_{\theta}(X)}-1\right)^{2}\right\rangle _{\theta}}\nonumber \\
 & =\frac{\left[\left\langle \Theta(X)\right\rangle _{\vartheta}-\left\langle \Theta(X)\right\rangle _{\theta}\right]^{2}}{\mathbb{D}_{\mathrm{PE}}\left[P_{\vartheta}||P_{\theta}\right]},\label{eq:AppA_ChapmanRobbins_ineq}
\end{align}
where $\mathbb{D}_{\mathrm{PE}}\left[P_{\vartheta}||P_{\theta}\right]$
is the Pearson divergence: 
\begin{align}
\mathbb{D}_{\mathrm{PE}}\left[P_{\vartheta}||P_{\theta}\right] & \equiv\int dX\,\left(\frac{P_{\vartheta}(X)}{P_{\theta}(X)}-1\right)^{2}P_{\theta}(X)\nonumber \\
 & =\int dX\,\left(\frac{P_{\vartheta}(X)}{P_{\theta}(X)}\right)^{2}P_{\theta}(X)-1.\label{eq:Pearson_def}
\end{align}

\section{Path integral\label{sec:path_integral}}

Here we introduce the pre-point discretization procedure of the path
integral according to Refs.~\cite{Wio:2013:PIBook,Bressloff:2014:WKB}.
We focus on the one-dimensional case because
the calculations for the
multi-dimensional case is
laborious.

We consider the following Langevin equation (Ito interpretation) 
\begin{equation}
\dot{x}=A_{\theta}(x,t)+\sqrt{2}C(x,t)\xi(t).\label{eq:Ito_Langevin_def}
\end{equation}
We discretize time by dividing the interval $[0,T]$ into $K$ equipartitioned
intervals with time resolution $\Delta t$, where $T=K\Delta t$,
$t^{k}\equiv k\Delta t$, and $x^{k}\equiv x(t^{k})$ (superscripts
denote points in a temporal sequence). Discretization of Eq.~\eqref{eq:Ito_Langevin_def}
yields 
\begin{equation}
x^{k+1}-x^{k}=\Delta tA_{\theta}(x^{k},t^{k})+\sqrt{2}C(x^{k},t^{k})\Delta w^{k},\label{eq:discretized_Langevin}
\end{equation}
where $\Delta w^{k}\equiv w^{k+1}-w^{k}=w(t^{k+1})-w(t^{k})$ with
$w(t)$ depicting the Wiener process. $\Delta w^{k}$ has the
following properties 
\begin{equation}
\left\langle \Delta w^{k}\right\rangle =0,\hspace*{1em}\left\langle \Delta w^{k}\Delta w^{k^{\prime}}\right\rangle =\Delta t\delta_{kk'}.\label{eq:AppB_Wiener_prop}
\end{equation}
A stochastic trajectory $\mathcal{X}\equiv\left[x^{1},x^{2},...,x^{K}\right]$
is specified, given $\mathcal{W}\equiv\left[\Delta w^{0},\Delta w^{1},...,\Delta w^{K-1}\right]$
and $x^{0}$. The Wiener process $\Delta w^{k}$ has the following
probability density function: 
\begin{equation}
P(\mathcal{W})=\prod_{k=0}^{K-1}P(\Delta w^{k})=\prod_{k=0}^{K-1}\frac{1}{\sqrt{2\pi\Delta t}}\exp\left[-\frac{(\Delta w^{k})^{2}}{2\Delta t}\right].\label{eq:w_PDF_def}
\end{equation}
Let us change variables in Eq.~\eqref{eq:w_PDF_def} from $\mathcal{W}=\left[\Delta w^{0},\Delta w^{1},...,\Delta w^{K-1}\right]$
to $\mathcal{X}=\left[x^{1},x^{2},...,x^{K}\right]$. From Eq.~\eqref{eq:discretized_Langevin},
the determinant of a Jacobian matrix is 
\begin{equation}
\left|\frac{\partial(x^{1},...,x^{K})}{\partial(\Delta w^{0},...,\Delta w^{K-1})}\right|=\prod_{k=0}^{K-1}\sqrt{2B(x^{k},t^{k})},\label{eq:Jacobian_def}
\end{equation}
given that the determinant of a triangular matrix is a product of
its diagonal elements, where we used $B(x,t)\equiv C(x,t)^{2}$.
Using Eqs.~\eqref{eq:discretized_Langevin}, \eqref{eq:w_PDF_def},
and \eqref{eq:Jacobian_def}, we obtain \footnote{Instead of using the Jacobian, Ref.~\cite{Bressloff:2014:WKB} used
a $\delta$ function to change the variables. The two approaches are
equivalent. } \begin{widetext}
\begin{equation}
P_{\theta}(\mathcal{X}|x^{0})=\left(\prod_{k=0}^{K-1}\frac{1}{\sqrt{4\pi\Delta tB(x^{k},t^{k})}}\right)\exp\left[-\frac{1}{4}\sum_{k=0}^{K-1}\Delta t\left\{ \left(\frac{x^{k+1}-x^{k}}{\Delta t}-A_{\theta}(x^{k},t^{k})\right)^{2}B(x^{k},t^{k})^{-1}\right\} \right].\label{eq:PX_pathintegral_def}
\end{equation}
\end{widetext}In the limit $K\rightarrow\infty$, $\mathcal{X}\to\Gamma\equiv[x(t)]_{t=0}^{t=T}$
and we can write 
\begin{equation}
\mathcal{P}_{\theta}(\Gamma|x^{0})=\mathscr{N}\exp\left[-\frac{1}{4}\int_{0}^{T}dt\,\left(\dot{x}-A_{\theta}(x,t)\right)^{2}B(x,t)^{-1}\right].\label{eq:AppB_pathprob}
\end{equation}
For an arbitrary function $g(x,t)$, the following relation holds
\begin{align}
 & \left\langle \int_{0}^{T}dt\,\left(\dot{x}-A_{\theta}(x,t)\right)\bullet g(x,t)\right\rangle _{\theta}\nonumber \\
 & =\left\langle \sum_{k=0}^{K-1}\Delta t\left\{ \left(\frac{x^{k+1}-x^{k}}{\Delta t}-A_{\theta}(x^{k},t^{k})\right)g(x^{k},t^{k})\right\} \right\rangle _{\theta}\nonumber \\
 & =\left\langle \sum_{k=0}^{K-1}\sqrt{2}\Delta w^{k}C(x^{k},t^{k})g(x^{k},t^{k})\right\rangle _{\theta}=0,\label{eq:non_anticipating_vanish}
\end{align}
where we used the property that $x^k$ does not depend on $\Delta w^k$. 

\section{Ito and Stratonovich currents\label{sec:currents}}

We present a relation between Ito and Stratonovich currents {[}cf.
Eq.~\eqref{eq:Ito_Str_currents_def}{]}, both of which appear in
the main text. Here, we explain this relation for the one-dimensional case,
with the multi-dimensional generalization presented later.

Ito and its equivalent Stratonovich Langevin equations are given by
\begin{align}
dx & =A(x)dt+\sqrt{2}C(x)\bullet dw,\label{eq:Ito_def}\\
dx & =\left[A(x)-C(x)C^{\prime}(x)\right]dt+\sqrt{2}C(x)\circ dw,\label{eq:Stratonovich_def}
\end{align}
respectively. Let $\eta(x)$ be an arbitrary function of $x$. We
are concerned with the relation between the following two terms: 
\begin{equation}
U_{I}\equiv\int_{0}^{T}\eta(x)\bullet dw,\hspace*{1em}U_{S}\equiv\int_{0}^{T}\eta(x)\circ dw.\label{eq:UI_US_def}
\end{equation}
Their discretized representations are 
\begin{align}
U_{I} & =\sum_{k=0}^{K-1}\eta\left(x^{k}\right)\Delta w^{k},\label{eq:UI_def}\\
U_{S} & =\sum_{k=0}^{K-1}\eta\left(\frac{x^{k+1}+x^{k}}{2}\right)\Delta w^{k}.\label{eq:US_def}
\end{align}
Applying a Taylor series expansion to Eq.~\eqref{eq:US_def} and
dropping terms whose orders are higher than $O(\Delta t)$, we obtain
the following well-known relation \cite{Gardiner:2009:Book}:
\begin{align}
U_{S} & =\sum_{k=0}^{K-1}\left[\eta(x^{k})\Delta w^{k}+\frac{\sqrt{2}}{2}C(x^{k})\eta^{\prime}(x^{k})(\Delta w^{k})^{2}\right]\nonumber \\
 & =\int_{0}^{T}\eta(x)\bullet dw+\frac{\sqrt{2}}{2}\int_{0}^{T}C(x)\eta^{\prime}(x)\bullet dw^{2}\nonumber \\
 & =U_{I}+\frac{\sqrt{2}}{2}\int_{0}^{T}C(x)\eta^{\prime}(x)dt,\label{eq:US_UI_relation}
\end{align}
where we used the relation $dw^{2}=dt$ in the last line, which is valid
for any non-anticipating function (see Chapter 4 in \cite{Gardiner:2009:Book}
for details).

Next, we consider Ito and Stratonovich currents of the following forms:
\begin{equation}
J_{I}\equiv\int_{0}^{T}\Lambda(x)\bullet\dot{x}dt,\hspace*{1em}J_{S}\equiv\int_{0}^{T}\Lambda(x)\circ\dot{x}dt,\label{eq:Ito_Str_currents_def}
\end{equation}
where $\Lambda(x)$ is a projection function. Their discretized representations
are 
\begin{align}
J_{I} & =\sum_{k=0}^{K-1}\Lambda\left(x^{k}\right)\left(x^{k+1}-x^{k}\right).\label{eq:JI_def}\\
J_{S} & =\sum_{k=0}^{K-1}\Lambda\left(\frac{x^{k+1}+x^{k}}{2}\right)\left(x^{k+1}-x^{k}\right),\label{eq:JS_def}
\end{align}
Substituting Eqs.~\eqref{eq:Ito_def} and \eqref{eq:Stratonovich_def}
into Eqs.~\eqref{eq:JI_def} and \eqref{eq:JS_def}, respectively,
we obtain 
\begin{align}
J_{I} & =\int_{0}^{T}\Lambda(x)A(x)dt+\sqrt{2}\int_{0}^{T}\Lambda(x)C(x)\bullet dw.\label{eq:JI2}\\
J_{S} & =\int_{0}^{T}\Lambda(x)\left(A(x)-C(x)C^{\prime}(x)\right)dt\nonumber \\
 & +\sqrt{2}\int_{0}^{T}\Lambda(x)C(x)\circ dw,\label{eq:JS2}
\end{align}
By using Eq.~\eqref{eq:US_UI_relation} {[}$\eta(x)=\Lambda(x)C(x)${]},
the following relation holds: 
\begin{align}
\int_{0}^{T}\Lambda(x)C(x)\circ dw & =\int_{0}^{T}\Lambda(x)C(x)\bullet dw\nonumber \\
 & +\frac{\sqrt{2}}{2}\int_{0}^{T}C(x)\frac{d\Lambda(x)C(x)}{dx}dt.\label{eq:JS2_term}
\end{align}
By substituting Eq.~\eqref{eq:JS2_term} into Eq.~\eqref{eq:JS2},
a relation between the Stratonovich current $J_{S}$ and the Ito current
$J_{I}$ is given by 
\begin{align}
J_{S} & =\int_{0}^{T}\Lambda(x)\left(A(x)-C(x)C^{\prime}(x)\right)dt\nonumber \\
 & +\sqrt{2}\left[\int_{0}^{T}\Lambda(x)C(x)\bullet dw+\frac{\sqrt{2}}{2}\int_{0}^{T}C(x)\frac{d\Lambda(x)C(x)}{dx}dt\right]\nonumber \\
 & =\int_{0}^{T}\Lambda(x)A(x)dt+\sqrt{2}\int_{0}^{T}\Lambda(x)C(x)\bullet dw\nonumber \\
 & +\int_{0}^{T}\Lambda^{\prime}(x)C(x)^{2}dt\nonumber \\
 & =J_{I}+\int_{0}^{T}\Lambda^{\prime}(x)C(x)^{2}dt.\label{eq:JS_JI_diff}
\end{align}
Therefore, we find the following relation:
\begin{equation}
\int_{0}^{T}\Lambda(x)\circ\dot{x}dt=\int_{0}^{T}\Lambda(x)\bullet\dot{x}dt+\int_{0}^{T}B(x)\frac{d\Lambda(x)}{dx}dt.\label{eq:JS_JI_result}
\end{equation}
For the multi-dimensional case $d\bm{x}=\bm{A}(\bm{x})dt+\sqrt{2}\bm{C}(\bm{x})\bullet d\bm{w}$,
we repeat the same calculations to obtain 
\begin{align}
\int_{0}^{T}\bm{\Lambda}(\bm{x})^{\top}\circ\dot{\bm{x}}dt & =\int_{0}^{T}\bm{\Lambda}(\bm{x})^{\top}\bullet\dot{\bm{x}}dt\nonumber \\
 & +\int_{0}^{T}\mathrm{Tr}\left[\bm{B}(\bm{x})\frac{\partial\bm{\Lambda}(\bm{x})}{\partial\bm{x}}\right]dt,\label{eq:JS_JI_mult_result}
\end{align}
where $\bm{B}(\bm{x})\equiv\bm{C}(\bm{x})\bm{C}(\bm{x})^{\top}$ and
$\left[\partial\bm{\Lambda}(\bm{x})/\partial\bm{x}\right]_{ij}\equiv\partial\Lambda_{i}(\bm{x})/\partial x_{j}$
is a Jacobian matrix.

\section{Steady-state distribution of periodic systems\label{sec:periodic_potential}}

Because $\Theta_{\mathrm{tot}}(\Gamma)$ is defined through the projection
function $\Lambda(x)=1/[P^{\mathrm{ss}}(x)B(x)]$ {[}Eq.~\eqref{eq:Theta_tot_def}{]},
we need to calculate the steady-state distribution $P^{\mathrm{ss}}(x)$,
which can be found analytically as shown below.

Let $f(x)$ be a periodic potential $f(x)=f(x+2\pi)$ and $\rho\ge0$
be an external force. Suppose a system is given by 
\begin{equation}
\dot{x}=\rho-f^{\prime}(x)+\sqrt{2D}\xi(t),\label{eq:periodic_system_def}
\end{equation}
where $\rho=ab$ and $f(x)=-\frac{1}{2}(a+\sin(x))^{2}+b\cos(x))$
for Eq.~\eqref{eq:Ax_periodic_def}. According to Ref.~\cite{Risken:1989:FPEBook}
(p.287), the steady-state distribution of Eq.~\eqref{eq:periodic_system_def}
is 
\begin{equation}
P^{\mathrm{ss}}(x)=\exp\left(-\frac{V(x)}{D}\right)\left[\mathfrak{N}-\frac{J^{\mathrm{ss}}}{D}\int_{0}^{x}\exp\left(\frac{V(x^{\prime})}{D}dx^{\prime}\right)\right],\label{eq:AppD_Pss}
\end{equation}
where $V(x)=f(x)-\rho x$ and $\mathfrak{N}$ is a normalization constant.
$\mathfrak{N}$ and $J^{\mathrm{ss}}$ are determined by the two constraints:
\begin{align}
J^{\mathrm{ss}}\int_{0}^{2\pi}\exp\left(\frac{V(x)}{D}\right)dx & =D\mathfrak{N}\left[1-\exp\left(-\frac{2\pi\rho}{D}\right)\right],\label{eq:periodic_constraint1}\\
\int_{0}^{2\pi}P^{\mathrm{ss}}(x)dx & =1.\label{eq:periodic_constraint2}
\end{align}
Equations~\eqref{eq:periodic_constraint1} and \eqref{eq:periodic_constraint2}
are solved numerically to obtain $\mathfrak{N}$ and $J^{\mathrm{ss}}$.

\section{Pearson divergence for linear Langevin process\label{sec:OU_bound}}

The Pearson divergence is calculated analytically for a linear Langevin
equation of Eq.~\eqref{eq:A_linear_def} (the Ornstein--Uhlenbeck
process). The discretized representation of Eq.~\eqref{eq:A_linear_def}
is
\begin{equation}
x^{k+1}-x^{k}=\Delta x^{k}=\left[-\alpha x^{k}+\theta u^{k}\right]\Delta t+\sqrt{2D}\Delta w^{k},\label{eq:AppE_discrete_Langevin}
\end{equation}
where $u^{k}\equiv u(t^{k})$. The probability of the discretized
trajectory $\mathcal{X}=[x^{1},x^{2},...,x^{K}]$ given $x^{0}$ is
{[}cf. Eq.~\eqref{eq:PX_pathintegral_def}{]} 
\begin{align}
 & P_{\theta}(\mathcal{X}|x^{0})=\frac{1}{(4\pi D\Delta t)^{K/2}}\nonumber\\
 & \times\exp\left[-\frac{1}{4D\Delta t}\sum_{k=0}^{K-1}\left(x^{k+1}-x^{k}-\{-\alpha x^{k}+\theta u^{k}\}\Delta t\right)^{2}\right].\label{eq:AppE_pathprob_1}
\end{align}
The Pearson divergence between $P_{\theta}(\mathcal{X})$ and $P_{\theta=0}(\mathcal{X})$
is {[}cf. Eq.~\eqref{eq:Pearson_def}{]} 
\begin{align}
\mathbb{D}_{\mathrm{PE}}\left[P_{\theta}||P_{\theta=0}\right] & =\int\prod_{k=0}^{K}dx^{k}\left[\frac{P_{\theta}(\mathcal{X}|x^{0})P_{\theta}(x^{0})}{P_{\theta=0}(\mathcal{X}|x^{0})P_{\theta=0}(x^{0})}\right]^{2}\nonumber\\
 & \times P_{\theta=0}(\mathcal{X}|x^{0})P_{\theta=0}(x^{0})-1.\label{eq:AppE_DPE}
\end{align}
Let us introduce new variables $y^{k}$ ($k=1,2,...,K$), defined
by 
\begin{equation}
y^{k+1}\equiv x^{k+1}-x^{k}+\alpha x^{k}\Delta t.\label{eq:yk_def}
\end{equation}
The determinant of a Jacobian is 
\begin{equation}
\left|\frac{\partial(y^{1},y^{2},...,y^{K})}{\partial(x^{1},x^{2},...,x^{K})}\right|=1.
\label{eq:yx_Jacobian_det}
\end{equation}
Using Eqs.~(\ref{eq:yk_def}) and (\ref{eq:yx_Jacobian_det}), the
probability density of $\mathcal{Y}\equiv[y^{1},y^{2},...,y^{K}]$
is 
\begin{align}
P_{\theta}(\mathcal{Y}|x^{0}) & =\frac{1}{(4\pi D\Delta t)^{K/2}}\exp\left[-\frac{1}{4D\Delta t}\sum_{k=0}^{K-1}\left(y^{k+1}-\theta u^{k}\Delta t\right)^{2}\right].\label{eq:AppE_pathprob_Y}
\end{align}
Therefore, the Pearson divergence is given by \begin{widetext}
\begin{align}
\mathbb{D}_{\mathrm{PE}}\left[P_{\theta}||P_{\theta=0}\right] & =\int dx^{0}\int\prod_{k=1}^{K}dy^{k}\left[\frac{P_{\theta}(\mathcal{Y}|x^{0})P_{\theta}(x^{0})}{P_{\theta=0}(\mathcal{Y}|x^{0})P_{\theta=0}(x^{0})}\right]^{2}P_{\theta=0}(\mathcal{Y}|x^{0})P_{\theta=0}(x^{0})-1\nonumber\\
 & =-1+\int dx^{0}\left(\frac{P_{\theta}(x^{0})}{P_{\theta=0}(x^{0})}\right)^{2}P_{\theta=0}(x^{0})\int\prod_{k=1}^{K}\frac{dy^{k}}{(4\pi D\Delta t)^{K/2}}\nonumber\\
 & \times\exp\left[-\frac{1}{2D\Delta t}\sum_{k=0}^{K-1}\left(y^{k+1}-\theta u^{k}\Delta t\right)^{2}+\frac{1}{2D\Delta t}\sum_{k=0}^{K-1}\left(y^{k+1}\right)^{2}-\frac{1}{4D\Delta t}\sum_{k=0}^{K-1}\left(y^{k+1}\right)^{2}\right]\nonumber\\
 & =-1+\exp\left[\sum_{k=0}^{K-1}\frac{(u^{k})^{2}\Delta t}{2D}\theta^{2}\right]\int dx^{0}\left(\frac{P_{\theta}(x^{0})}{P_{\theta=0}(x^{0})}\right)^{2}P_{\theta=0}(x^{0}).\label{eq:AppE_DPE_final}
\end{align}
\end{widetext}When the initial distributions are the same for $\theta\ne0$
and $\theta=0$, in the limit of $K\to\infty$, we obtain Eq.~\eqref{eq:DPE_OU}. 

\section{Numerical simulations\label{sec:simulations}}

In the main text, we performed numerical simulations. In this section,
we explain how these simulations are implemented.

\subsection{Monte Carlo simulations}

In order to solve the Ito Langevin equations, we used Eq.~\eqref{eq:discretized_Langevin}
(this method is known as the Euler--Maruyama scheme). Stratonovich-type
currents are calculated by Eq.~\eqref{eq:JS_def}.

We numerically calculate the Pearson divergence. We generate trajectories
from the Langevin equations with parameter $\theta=0$. Let $N_{S}$
be the number of generated trajectories and $\bm{\mathcal{X}}_{i}$
be the $i$th realization of the trajectories. Let us consider $\bm{B}(\bm{x},t)=\bm{D}$.
The integral of the Pearson divergence is approximated by the following
summation: 
\begin{equation}
\mathbb{D}_{\mathrm{PE}}\left[\mathcal{P}_{\theta}||\mathcal{P}_{\theta=0}\right]\simeq\frac{1}{N_{S}}\sum_{i=1}^{N_{S}}\left(\frac{P_{\theta}(\bm{\mathcal{X}}_{i})}{P_{\theta=0}(\bm{\mathcal{X}}_{i})}-1\right)^{2},\label{eq:PE_discrete}
\end{equation}
where\begin{widetext}
\begin{align}
P_{\theta}(\bm{\mathcal{X}}|\bm{x}_{0})=\exp\left[-\frac{\Delta t}{4}\sum_{k=0}^{K-1}\sum_{i,j}\left(\frac{x_{i}^{k+1}-x_{i}^{k}}{\Delta t}-A_{\theta,i}(\bm{x}^{k},t^{k})\right)D_{ij}^{-1}\left(\frac{x_{j}^{k+1}-x_{j}^{k}}{\Delta t}-A_{\theta,j}(\bm{x}^{k},t^{k})\right)\right].\label{eq:AppF_pathprob_X}
\end{align}
\end{widetext}Here $D_{ij}^{-1}$ is an $i,j$th element of $\bm{D}^{-1}$.
We omitted $\mathscr{N}$ because it cancels out in Eq.~\eqref{eq:PE_discrete}.

\subsection{Phase definition}

The phase for limit cycle oscillators can be defined \cite{Kuramoto:2003:OscBook}.
For deterministic oscillators, we define phase $\phi$ on a closed
orbit by 
\begin{equation}
\frac{d\phi}{dt}=\Omega,\label{eq:phase_def_1}
\end{equation}
where $\Omega$ is the angular frequency of the deterministic oscillation.
We can expand the definition of the phase into an entire space $\bm{x}$,
where $\bm{x}$ is an $N$-dimensional vector.
Let $\bm{x}_{a}$ be a point on the closed orbit and $\bm{x}_{b}$
be a point that is \emph{not} on the orbit. According to Eq.~\eqref{eq:phase_def_1},
we can determine $\phi(\bm{x}_{a})$. As the closed orbit is an attractor
in limit cycle oscillators, $\bm{x}_{b}$ eventually converge to the
closed orbit for $t\to\infty$. We let $\bm{x}_{a}$ and $\bm{x}_{b}$
time-evolve for the same duration. If the two points eventually converge
to the same point on the closed orbit, then we assign $\phi(\bm{x}_{b})=\phi(\bm{x}_{a})$.

\end{document}